\documentclass[12pt]{article}
\usepackage{amsmath}
\usepackage{amssymb}
\usepackage{latexsym}
\usepackage{epsfig}

\newcommand{\be}{\begin{equation}}
\newcommand{\ee}{\end{equation}}

\begin{document}
\begin{center}
\large \textbf{Thermodynamics and weak cosmic censorship conjecture in Reissner-Nordstr$\ddot{o}$m anti-de Sitter black holes with scalar field}
\end{center}

\begin{center}
Deyou Chen $^{a}$\footnote{ E-mail: {deyouchen@hotmail.com}},
Wei Yang $^{b}$\footnote{E-mail: {yangweicdu@hotmail.com}},
Xiaoxiong Zeng $^{c}$\footnote{E-mail: {xxzengphysics@163.com}}
\end{center}

\begin{center}

$^{a}$ School of Science, Xihua University, Chengdu 610039, China\\
$^{b}$ School of Information Science and Engineering, Chengdu University, Chengdu 610106, China\\
$^{c}$ School of Material Science and Engineering, Chongqing Jiaotong University, Chongqing 400074, China
\end{center}

{\bf Abstract:} The thermodynamics and weak cosmic censorship conjecture in Reissner-Nordstr$\ddot{o}$m anti-de Sitter black holes are investigated by the scattering of the scalar field. The first law of thermodynamics in the non-extremal Reissner-Nordstr$\ddot{o}$m anti-de Sitter black hole is recovered by the scattering. The increase of the horizon radius indicates that the singularity is not naked in this black hole. For the near-extremal and extremal black holes, the validity is tested by the minimum values of the function $f$ at their final states. It is found that both of the near-extremal and extremal black holes can not be overcharged. When $\omega=q\phi$, the final state of the extremal black hole is still an extremal black hole. When $\omega\neq q\phi$, it becomes a near-extremal black hole with new mass and charge.

\section{Introduction}

It is widely believed that spacetime singularities arise from gravitational collapse. In the vicinity of singularities, curvatures of spacetimes tend to diverge, and physical laws break down. To avoid these phenomena, Penrose proposed the weak cosmic censorship conjecture (WCCC) in 1969 \cite{RP}. It is stated in the conjecture that naked singularities cannot be produced in a real physical process from regular initial conditions and indicates that singularities are hidden behind event horizons without any access to distant observers. Although this conjecture seems to be reasonable, a concrete proof is lacked and people can only test its validity.

The Gedanken experiment is an effective method to test the WCCC. In this experiment, a test particle with energy and large enough charge and angular momentum is thrown into a black hole to see whether its event horizon is destroyed or not after the particle is absorbed. If the horizon is destroyed, the singularity is naked; on the contrary, the singularity is hidden behind the horizon. The destruction of the horizon is judged by the disappearance of the horizon solution. For a Kerr-Newman black hole, the existence of the event horizon satisfies the condition $M^2 \ge a^2 + Q^2$. When $M^2 < a^2 + Q^2$, the horizon disappears and the naked singularity appears. Thus, the WCCC is violated. Wald is a pioneer deviser of this experiment  \cite{RMW}. His research showed the test particle could not be captured by the extremal Kerr-Newman black hole and the naked singularity did not exist. Since this seminal experiment was proposed, people have studied the validity of the WCCC in the various spacetimes. Hubeny found that a near-extremal Reissner-Nordstr$\ddot{o}$m black hole would be overcharged by absorbing a test particle with energy and large enough charge \cite{VEH}. This result was obtained again in the work of Matsas and Silva, where the tunnelling effects were taken into account \cite{MS}. Similarly, when a particle with energy and large enough angular momentum was absorbed by a rotating black hole, the black hole could be overspun \cite{SH1,JS,RC,SS,GZ,RS1}. However, when the backreaction and self-force effects were taken into account \cite{SH2,BCK1,BCK2,ZVPH,IST,CB,CBSM}, the particles might be not captured by the black holes and the singularises could not be naked.

The validity of the WCCC can be studied through the Gedanken experiment by replacing a test particle with a test field. This method was first developed by Semiz \cite{IS}. Initially, there is no field, only a black hole. The test field comes in from infinity and interacts with the black hole. The field is required to have a finite energy. Part of the field is absorbed by the black hole, and the rest is reflected back to infinity. Energy, charge and angular momentum are transferred between the field and the black hole. Finally, the field decays away leaving behind a spacetime with the new energy, charge and angular momentum. The validity is determined by whether the event horizon exists in the new spacetime. Considering the interaction between a complex scalar field and a charged black hole, Semiz found that the WCCC was not violated in the dyonic Kerr-Newman black holes. This result was gotten again in \cite{GZT}. But it is not consistent with those gotten in the Kerr and BTZ black holes and in the Dirac field \cite{DS1,KD,DS2}. Based on this experiment, Gwak recently researched on the validity of the WCCC in the Kerr (anti-)de Sitter black holes by the scattering of a scalar field \cite{BG3}. He found that the black holes were not overspun. In his work, the energy and angular momentum in a certain time interval were calculated by their fluxes, respectively. The time interval was chosen as infinitesimal, which leads to the infinitesimal transfer of energy and angular momentum. Therefore, the self-force effect and other interactions could be neglected in the scattering process. This result supports the WCCC and is in agreement with that derived by Wald and Sorce \cite{RMW2,SW}. Other researches on the validity of the WCCC are referred to \cite{MST,CO,LCNR,RS,SH3,NQV,HS,RV,BG1,BG2,CHS2,GMZZ,KD2,KD3,CHS,ASZZ,YW,LWL,DJ,GG,BG5,KD5,CHEN} and the references therein. Although a lot of work has been done on the validity of the WCCC, there is still no identical conclusion.

In this paper, we investigate the thermodynamics and WCCC in Reissner-Nordstr$\ddot{o}$m anti-de Sitter (RN-AdS) black holes by the scattering of a complex scalar field. Due to the existence of the electromagnetic filed, the interaction between the charge and the black hole is taken into account. The derivation of the transferred charge relies on the charge flux and a certain time interval. Here the time interval is infinitesimal, and then the self-force effect is neglected. The energy is gotten by its flux determined by the energy-momentum tensor. For the non-extremal RN-AdS black hole, the first law of thermodynamics is recovered by the scattering of the scalar field. The increase of the entropy supports the second law of thermodynamics. The validity of the WCCC in this black hole is verified by the increase of the horizon radius. For the near-extremal and extremal RN-AdS black holes, due to the divergence of Eq.(\ref{eq3.8}), the validity is tested by the minimum value of the function $f$ at the final state after the scattering. This value is evaluated by adjusting values of $\omega$ and $q$. If the value is negative, the event horizon exists and the singularity is not naked. Our result shows that both of the near-extremal and extremal RN-AdS black holes can not be overcharged.

The rest of this paper is organized as follows. In the next section, we solve the ingoing wave function in the RN-AdS black holes by using the field equation and introducing the tortoise coordinate. In section 3, the thermodynamics of the non-extremal RN-AdS black hole is discussed by the scattering of the scalar field. The WCCC  is tested by the change of the horizon radius. In section 4, we investigate the validity of the WCCC in the near-extremal and extremal RN-AdS black holes by the minimum values of the function $f$. Section 5 is devoted to our discussion and conclusion. Throughout the paper, we set $G = c = \hbar = 1$.

\section{The wave function in the RN-AdS black hole}

The purpose of this section is to solve the ingoing wave function in the RN-AdS black hole. The metric of the RN-AdS black hole is given by

\begin{eqnarray}
ds^2 = -fdt^2 + \frac{1}{f}dr^2 + r^2 (d\theta^2 + sin^2\theta d\varphi^2),
\label{eq2.1}
\end{eqnarray}

\noindent with the electromagnetic potential

\begin{eqnarray}
A_{\mu}=\left(-\frac{Q}{r},0,0,0\right),
\label{eq2.2}
\end{eqnarray}

\noindent where $f=f(M,Q,r)=1-\frac{2M}{r}+\frac{Q^2}{r^2}+\frac{r^2}{l^2}$, $M$ and $Q$ are the physical mass and charge, respectively. $l^2$ is a constant related to the cosmological constant as $\Lambda=-3l^{-2}$. In general case, $f$ has two positive real roots, which corresponds to the Cauchy horizon $r_-$ and the event horizon $r_+$, respectively. When these two horizons coincide with each other, the black hole is extremal and its Hawking temperature is zero. The mass relates to its horizon radius $r_+$ as $2M= r_++\frac{Q^2}{r_+}+\frac{r_+^3}{l^2}$. The black hole entropy is $S=\pi r_+^2$. The Hawking temperature is

\begin{eqnarray}
T=\left. \frac{1}{4\pi}\frac{\partial f(M,Q,r)}{\partial r}\right|_{r=r_+} =\frac{1}{2\pi}\left(\frac{M}{r_+^2}-\frac{Q^2}{r_+^3}+\frac{r_+}{l^2}\right).
\label{eq2.3}
\end{eqnarray}

The action of the complex scalar field in the fixed RN-AdS gravitational and electromagnetic fields is

\begin{eqnarray}
\mathcal{S} &=& \int{\sqrt{-g}\mathcal{L}d^4x} \nonumber\\
&=& -\frac{1}{2}\int{\sqrt{-g}[(\partial^{\mu}-ieA^{\mu})\Psi^{*}(\partial_{\mu}+ieA_{\mu})\Psi+m^2\Psi^{*}\Psi]d^4x},
\label{eq2.4}
\end{eqnarray}

\noindent where $\mathcal{L}$ is the Lagrangian density, $A_{\mu}$ is the electromagnetic potential given by Eq.(\ref{eq2.2}), $m$ is the mass, $\Psi$ denotes the wave function and its conjugate is $\Psi^*$. The field equation gotten from the action satisfies

\begin{eqnarray}
\left(\nabla ^{\mu}-ieA^{\mu}\right)\left(\nabla _{\mu}-ieA_{\mu}\right)\Psi -m^2\Psi=0.
\label{eq2.5}
\end{eqnarray}

\noindent The wave function $\Psi$ in the RN-AdS spacetime is solved from the field equation. To solve this wave function, we carry out a separation of variables

\begin{eqnarray}
\Psi= e^{-i\omega t}R(r)\Theta(\theta ,\varphi).
\label{eq2.6}
\end{eqnarray}

\noindent In the above equation, $\omega$ is the energy of the particle, $\Theta(\theta ,\varphi)$ are the scalar spherical harmonics. We insert the contravariant components of the RN-AdS metric and the wave function into the field equation. Introducing a separation variable constant $\lambda$ yields two equations

\begin{eqnarray}
\frac{d}{dr}\left(r^2f\frac{d}{dr}R(r)\right)-\left[\lambda +m^2r^2-\frac{1}{r^2f}(\omega r^2-eQr)^2\right]R(r)= 0,
\label{eq2.7}
\end{eqnarray}

\begin{eqnarray}
\frac{1}{sin^2\theta} \frac{\partial^2\Theta(\theta,\varphi)}{\partial\varphi^2}+\frac{1}{sin\theta}\frac{\partial}{\partial \theta}\left(sin\theta\frac{\partial \Theta(\theta,\varphi)}{\partial \theta}\right) +\lambda\Theta(\theta,\varphi)= 0.
\label{eq2.7-1}
\end{eqnarray}

\noindent The separation constant takes on the form $\lambda = l(l+1)$ in which $l$ denotes an angular momentum number. The angular function $\Theta(\theta ,\varphi)$ is solved by Eq.(\ref{eq2.7-1}). It can be reduced to unity by the normalization condition in the calculation of the energy and charge fluxes in the next section. Thus we don't need to know the exact expression of $\Theta(\theta ,\varphi)$. We focus our attention on the radial equation. For the massless field, the radial equation was discussed in the derivation of the quasinormal modes \cite{WLM1,WLM2,BK}. To solve $R(r)$, we perform the tortoise coordinate transformation

\begin{eqnarray}
dr_* = \frac{1}{f}dr
\label{eq2.8}
\end{eqnarray}

\noindent on Eq.(\ref{eq2.7}) and get

\begin{eqnarray}
r^4\frac{d^2R}{dr_*^2}+2r^3f\frac{dR}{dr_*}-\left[(m^2r^2+\lambda)r^2f-(\omega r^2-eQr)^2\right]R=0.
\label{eq2.9}
\end{eqnarray}

\noindent Near the event horizon, $f\to 0$ and the above equation is reduced to

\begin{eqnarray}
\frac{d^2R}{dr_*^2}+ \left(\omega -\frac{e Q}{r_+}\right)^2 R=0.
\label{eq2.10}
\end{eqnarray}

\noindent At the event horizon, the electromagnetic potential is $\phi= \frac{Q}{r_+}$. The radial wave function is gotten as

\begin{eqnarray}
R=e^{\pm i(\omega - e\phi)r_*},
\label{eq2.11}
\end{eqnarray}

\noindent where $r_*$ is a function of $r$ and $+/-$ correspond to the solutions of the outgoing/ingoing radial waves. Thus the wave function is

\begin{eqnarray}
\Psi=e^{-i\omega t}e^{\pm i(\omega - e\phi)r_*}\Theta(\theta,\varphi).
\label{eq2.12}
\end{eqnarray}

\noindent Although the expression of $\Theta(\theta,\varphi)$ is not given in the above function, it does not affect our result. The thermodynamics and the validity of the WCCC in the RN-AdS black holes are discussed by the scattering of the ingoing wave at the event horizon in this paper. So we focus our attention on the ingoing wave function.

\section{Thermodynamics of the non-extremal RN-AdS black hole}

The thermodynamics of the non-extremal RN-AdS black hole have been studied \cite{CYZ,PL,MKP1,MKP2,NTW}. In this section, we discuss its thermodynamics by the scattering of the ingoing wave at the event horizon. The first law of thermodynamics is presumed to be non-existent and recovered by the scattering. The self-force effect and other interactions are neglected in this paper, therefore, the transferred energy and charge must be very small.

The transfer of the energy in a certain time interval is related to the energy flux determined by the energy-momentum tensor. From the action, the energy-momentum tensor is gotten as follows

\begin{eqnarray}
T^{\mu}_{\nu}=\frac{1}{2}[(\partial ^{\mu}-ieA^{\mu})\Psi^* \partial_{\nu}\Psi + (\partial ^{\mu}+ieA^{\mu})\Psi\partial_{\nu} \Psi^{*}]+\delta_{\nu}^{\mu}\mathcal{L}.
\label{eq3.1}
\end{eqnarray}

\noindent Combining the ingoing wave function and its conjugate with the energy-momentum tensor yields the energy flux

\begin{eqnarray}
\frac{dM}{dt}=\int{T^{r}_{t}\sqrt{-g}d\theta d\varphi}=\omega(\omega - e\phi)r_+^2\int{\Theta^2(\theta,\varphi)\sin\theta d\theta d\varphi}=\omega(\omega - e\phi)r_+^2.
\label{eq3.3}
\end{eqnarray}

\noindent The derivation of the last equal sign in the above equation depends on the normalization condition, $\int{\Theta^2(\theta,\varphi)\sin\theta d\theta d\varphi}=1$. The transfer of the charge in a certain time interval is related to the electric current of the scalar field. The electric current is obtained from the action and it is

\begin{eqnarray}
j^{\mu}=\frac{\partial \mathcal{L}}{\partial A_{\mu}}=-\frac{1}{2}ie[\Psi^{*}(\partial ^{\mu}+ieA^{\mu})\Psi - \Psi(\partial ^{\mu}-ieA^{\mu})\Psi^{*}].
\label{eq3.2}
\end{eqnarray}

\noindent Introducing the ingoing wave function and its conjugate yields the charge flux

\begin{eqnarray}
\frac{dQ}{dt}=-\int{j^r\sqrt{-g}d\theta d\varphi}=e(\omega - e\phi)r_+^2\int{\Theta^2(\theta,\varphi)\sin\theta d\theta d\varphi}=e(\omega - e\phi)r_+^2.
\label{eq3.4}
\end{eqnarray}

\noindent Here, the normalization condition was used. Therefore, the transferred energy and charge within the certain time interval are

\begin{eqnarray}
dM=\omega(\omega - e\phi)r_+^2dt, \quad \quad  dQ=e(\omega - e\phi)r_+^2dt,
\label{eq3.5}
\end{eqnarray}

\noindent respectively. Due to the transferred energy and charge are very small, the time $dt$ must be also very small. The decreases of the energy and charge of the scalar field are equal to the increases of these of the black hole. The positive and negative values of $dM$ and $dQ$ lie on the relation between $\omega$ and $e\phi$. When $\omega > e\phi$, the values of $dM$ and $dQ$ are positive, which implies the increases of the energy and charge of the black hole during the scattering process. When $\omega = e\phi$, the black hole's energy and charge remain unchanged. When $\omega < e\phi$, the energy and charge flow out of the event horizon. This shows that the scattering extracts the energy and charge of the black hole and the superradiation occurs. In fact, the appearance of superradiation should satisfy that the boundary condition of the scalar field is in the asymptotic region. Here we follow the work of Gwak and focus on the infinitesimal change in the RN-AdS spacetime \cite{BG3}. Therefore, our discussion does not rely on the asymptotic boundary conditions. The discussion of the superradiation in the RN-AdS spacetime is referred to \cite{BGA}.

We assume that the final state of the non-extremal RN-AdS black hole is still a black hole. The initial state of the black hole is represented by $(M,Q,r_+)$, and the final state is represented by $(M+dM,Q+dQ,r_++dr_+)$, where $dM$, $dQ$ and $dr_+$ denote the increases of the black hole's energy, charge and radius, respectively. The variation of the radius can be obtained from the metric component $f$. The functions $f(M+dM,Q+dQ,r_++dr_+)$ and $f(M,Q,r_+)$  satisfy the following relation

\begin{eqnarray}
f(M+dM,Q+dQ,r_++dr_+)&=&f(M,Q,r_+)+\left.\frac{\partial f}{\partial M}\right|_{r=r_+}dM + \left.\frac{\partial f}{\partial Q}\right|_{r=r_+}dQ  \nonumber\\
&&+ \left.\frac{\partial f}{\partial r}\right|_{r=r_+}dr_+,
\label{eq3.6}
\end{eqnarray}

\noindent where

\begin{eqnarray}
\left.\frac{\partial f}{\partial M}\right|_{r=r_+}=-\frac{2}{r_+},\quad \left.\frac{\partial f}{\partial Q}\right|_{r=r_+}=\frac{2Q}{r_+^2}, \quad
\left.\frac{\partial f}{\partial r}\right|_{r=r_+} = \frac{2M}{r_+^2}-\frac{2Q^2}{r_+^3}+\frac{2r_+}{l^2}=4\pi T.
\label{eq3.7}
\end{eqnarray}

\noindent Because both of the initial and final states are black holes, we get $f(M+dM,Q+dQ,r_++dr_+)=f(M,Q,r_+)= 0$. Thus, the variation of the radius is

\begin{eqnarray}
dr_+=\frac{(\omega - e \phi)^2r_+}{2\pi T}dt.
\label{eq3.8}
\end{eqnarray}

\noindent Obviously, the black hole radius increases during the scattering process when $\omega \neq e \phi$, while it remains unchanged when $\omega = e\phi$. Therefore, the singularity is hidden behind the event horizon. The scattering can not destroy the event horizon and the WCCC is valid in the non-extremal RN-AdS black hole.

We further discuss its thermodynamics. From the relation between the entropy and radius, the variation of the entropy takes on the form

\begin{eqnarray}
dS=2\pi r_+dr_+=\frac{(\omega - e \phi)^2r_+^2}{T}dt\ge 0.
\label{eq3.9}
\end{eqnarray}

\noindent It shows that the entropy never decreases in the chronological direction. This result supports the second law of thermodynamics. Using Eq.(\ref{eq3.5}) and Eq.(\ref{eq3.9}), we get

\begin{eqnarray}
dM =TdS +\phi dQ,
\label{eq3.10}
\end{eqnarray}

\noindent which is the first law of thermodynamics of the RN-AdS black hole recovered by the scattering of the scalar field. As discussed above, the final state is assumed to be a black hole, therefore, the establishment of the first law is inevitable. For the near-extremal and extremal RN-AdS black holes, when they absorb energy and large enough charge, their final states may not be black holes, thus the laws of thermodynamics do not necessarily hold. When $T\to 0$, Eq.(\ref{eq3.8}) diverges. The above method can not be applied to the tests of the WCCC in the near-extremal and extremal RN-AdS black holes. We must resort to other methods.

\section{The WCCC in the near-extremal and extremal RN-AdS black holes}

We begin this section by testing the validity of the WCCC in the near-extremal RN-AdS black hole. When the charge absorbed by the black hole is large enough than the absorbed energy, the event horizon disappears. The black hole is overcharged and the WCCC is violated. Therefore, we check the existence of the event horizon after the scattering. A simply method to check this existence is to evaluate the solution of the metric component $f$. If the solution exists, the function $f$ must intersect with the $r-$axis in the graph $f-r$. In other words, the function $f$ with a minimum negative value guarantees the existence of the event horizon.

Near the extremal RN-AdS black hole, the minimum value is

\begin{eqnarray}
f(M,Q,r_0)=1-\frac{2M}{r_0}+\frac{Q^2}{r_0^2}+\frac{r_0^2}{l^2}<0,
\label{eq4.1}
\end{eqnarray}

\noindent where $r_0$ is the location corresponding to the minimum value. Let $|f(M,Q,r_0)| \ll 1$ to ensure the near-extremity. We use $(M,Q,r_0)$ to represent the initial state and $(M+dM,Q+dQ,r_0+dr_0)$ to represent the finial state. The initial state is a  RN-AdS black hole, but the final state is not necessarily a black hole. The function $f(M+dM,Q+dQ,r_0+dr_0)$ corresponding to the final state is expressed in term of the function $f(M,Q,r_0)$ corresponding to the initial state, which is

\begin{eqnarray}
f(M+dM,Q+dQ,r_0+dr_0)&=&f(M,Q,r_0)+\left.\frac{\partial f}{\partial M}\right|_{r=r_0} dM + \left.\frac{\partial f}{\partial Q}\right|_{r=r_0}dQ \nonumber\\
&&+ \left.\frac{\partial f}{\partial r}\right|_{r=r_0}dr_0,
\label{eq4.2}
\end{eqnarray}

\noindent where

\begin{eqnarray}
 \left.\frac{\partial f}{\partial M}\right|_{r=r_0}=-\frac{2}{r_0},\quad \quad  \left.\frac{\partial f}{\partial Q}\right|_{r=r_0}=\frac{2Q}{r_0^2},\quad \quad  \left.\frac{\partial f}{\partial r}\right|_{r=r_0} =0 .
\label{eq4.3}
\end{eqnarray}

\noindent Inserting Eq.(\ref{eq3.5}) and Eq.(\ref{eq4.3}) into Eq.(\ref{eq4.2}) yields

\begin{eqnarray}
f(M+dM,Q+dQ,r_0+dr_0) &=& -\left(\frac{q}{\omega}\right)^2 \frac{2\omega^2Q^2r_{+}dt}{r_0^2}+ 2\frac{q}{\omega} \frac{r_+^2+r_0r_+}{r_0^2}Q\omega^2dt \nonumber\\
&& -\frac{2r_+^2\omega^2dt}{r_0} + f(M,Q,r_0).
\label{eq4.4}
\end{eqnarray}

\noindent The above equation can be regarded as a quadratic function of $\frac{q}{\omega}$. If we can adjust the parameters $\omega$ and $q$ to make sure that the maximum value of the above function near $r_0$ is less than 0, then the corresponding values of other $\omega$ and $q$ near $r_0$ are also less than $0$. This implies that this function has a minimum negative value. When $\frac{q}{\omega}=\frac{r_++r_0}{2Q}$, the maximum value is

\begin{eqnarray}
f(M+dM,Q+dQ,r_0+dr_0)_{max} = \frac{1}{2r_0^2}\left[2\Delta_0 +\omega^2r_+(r_+-r_0)^2dt\right],
\label{eq4.5}
\end{eqnarray}

\noindent where $\Delta_0=(r_0^2-r_+^2)-2M(r_0-r_+)+\frac{r_0^4-r_+^4}{l^2}<0$, and the relation $Q^2=2Mr_+-r_+^2-\frac{r_+^4}{l^2}$ was used. To evaluate the value of the above equation, we let $r_+=r_0+\epsilon$, where $0<\epsilon\ll 1$. The near-extremal condition is satisfied. We express $\Delta_0$ in terms of $\epsilon$ and $r_+$ as the following relation

\begin{eqnarray}
\Delta_0&=&-2r_+\epsilon+2M\epsilon-\frac{4r_+^3}{l^2}\epsilon +\epsilon^2+\frac{6r_+^2\epsilon^2-4r_+\epsilon^3+\epsilon^4}{l^2},\nonumber\\
&=& -2r_+\epsilon+2M\epsilon-\frac{4r_+^3}{l^2}\epsilon<0.
\label{eq4.6}
\end{eqnarray}

\noindent The second equal sign in the above equation was gotten by omitting the higher order terms of $\epsilon$. $dt$ is an infinitesimal scale and is set $dt \sim \epsilon$. This was discussed in \cite{BG3}. Thus, the maximum value is reduced to

\begin{eqnarray}
f(M+dM,Q+dQ,r_0+dr_0)_{max}=\frac{1}{r_0^2}\left(-2r_+\epsilon+2M\epsilon-\frac{4r_+^3}{l^2}\epsilon\right).
\label{eq4.7}
\end{eqnarray}

\noindent Similarly, the higher-order term of $\epsilon$ was omitted in the above equation. Due to $-2r_+\epsilon+2M\epsilon-\frac{4r_+^3}{l^2}\epsilon<0$, the maximum value near $r_0$ is less than zero. This indicates that the event horizon exists in the finial state. The black hole can not be overcharged by the scattering of the scalar filed. Therefore, the WCCC is valid in the near-extremal RN-AdS black hole. Since the final state of the near-extremal black hole is still a black hole after the scattering, the first law of thermodynamics holds.

For the extremal RN-AdS black hole, we still use the above method to test the WCCC. Now the function $f$ has a minimum value equal to zero. $r_+$ is the location corresponding to this value. After the test field is scattered by the black hole, the function corresponding to the final state is also expressed as Eq.(\ref{eq4.4}). Using $r_0=r_+$ and $f(M,Q,r_+)=0$ yields

\begin{eqnarray}
f(M+dM,Q+dQ,r_++dr_+) = -2r_+(\omega-q\phi)^2dt .
\label{eq4.8}
\end{eqnarray}

\noindent When $\omega=q\phi$, we get $f(M+dM,Q+dQ,r_++dr_+)=0$, which shows that the scattering does not change the minimum value. This implies that the final state of the extremal RN-AdS black hole is still an extremal black hole. When $\omega\neq q\phi$, we derive $f(M+dM,Q+dQ,r_++dr_+)<0$ where two positive real roots exist. Because $dt$ is infinitesimal, $|f(M+dM,Q+dQ,r_++dr_+)| \ll 1$. This is similar to the initial state of the near-extremal black hole. It means that when $\omega\neq q\phi$, the final state of the extremal black hole becomes a new near-extremal black hole after the scattering. Therefore, the extremal black hole can not be overcharged. The WCCC is valid in the extremal RN-AdS black hole.

\section{Discussion and Conclusion}

In this paper, we investigated the thermodynamics and WCCC in the RN-AdS black holes by the scattering of the scalar field. The energy and charge in a infinitesimal time interval were calculated by the energy and charge fluxes, respectively. For the non-extremal RN-AdS black hole, the first law of thermodynamics was recovered by the scattering. The increase of the horizon radius confirms the validity of the WCCC in this black hole. The entropy does not decrease, which supports the second law of thermodynamics. For the near-extremal and extremal RN-AdS black holes, due to the divergence of Eq.(\ref{eq3.8}), we tested the WCCC by evaluating the minimum values of the function $f$ at the final states. These values were obtained by adjusting the values of $\omega$ and $q$. It was found that both of the near-extremal and extremal RN-AdS black holes can not be overcharged during the scattering process. When $\omega=q\phi$, the final state of the extremal RN-AdS black hole is still an extremal black hole. When $\omega\neq q\phi$, its final state becomes a near-extremal black hole with new mass and charge. Therefore, the WCCC is valid in both of the near-extremal and extremal RN-AdS black holes. This result is in consistence with that gotten by Wald and Sorce \cite{RMW2,SW}. In \cite{KD5}, D$\ddot{u}$ztas found that the perturbation of Kerr-Newman black holes by neutrino fields leads to a generic overspinning of the black hole. In this work, the second order terms $(\delta J)^2$ and $(\delta M)^2$ were retained. In \cite{RMW2,NQV}, the first order terms were retained in the analyses. Therefore, his result is different from these in this paper and in \cite{RMW2,NQV}.

Previous researches have shown that the self-force effect and other interactions would prevent the black holes from being overcharged and overspun \cite{SH2,BCK1,BCK2,ZVPH,IST,CB,CBSM}. In our investigation, due to the infinitesimal time interval, the self-force effect and other interactions were neglected, and the black hole can still not be overcharged.

\vspace*{2.0ex}
\noindent \textbf{Acknowledgments}

\noindent This work is supported by the National Natural Science Foundation of China (Grant no. 11875095), Sichuan Province Education Department Project (Grant No. 17ZA0294), and the Fund of the 1315 Project of Chengdu University.


\begin{thebibliography}{99}

\bibitem{RP}
R. Penrose, \emph{Gravitational collapse: the role of General Relativity}, \emph{Riv. Nuovo Cim.} \textbf{1} (1969) 252.

\bibitem{RMW}
R.M. Wald, \emph{Gedanken experiments to destroy a black hole}, \emph{Ann. Phys.} \textbf{82} (1974) 548.

\bibitem{VEH}
V.E. Hubeny, \emph{Overcharging a black hole and cosmic censorship}, \emph{Phys. Rev.} \textbf{D 59} (1999) 064013.

\bibitem{MS}
G.E.A. Matsas and A.R.R. da Silva, \emph{Overspinning a nearly extreme charged black hole via a quantum tunneling process}, \emph{Phys. Rev. Lett.} \textbf{99} (2007) 181301.

\bibitem{SH1}
S. Hod, \emph{Cosmic censorship, area theorem, and self-energy of particles}, \emph{Phys. Rev.} \textbf{D 66} (2002) 024016.

\bibitem{JS}
T. Jacobson and T.P. Sotiriou, \emph{Over-spinning a black hole with a test body}, \emph{Phys. Rev. Lett.} \textbf{103} (2009) 141101.

\bibitem{RC}
J.V. Rocha and V. Cardoso, \emph{Gravitational perturbation of the BTZ black hole induced by test particles and weak cosmic censorship in AdS spacetime}, \emph{Phys. Rev.} \textbf{D 83} (2011) 104037.

\bibitem{SS}
A. Saa and R. Santarelli, \emph{Destroying a near-extremal Kerr-Newman black hole}, \emph{Phys. Rev.} \textbf{D 84} (2011) 027501.

\bibitem{GZ}
S. Gao and Y. Zhang, \emph{Destroying extremal Kerr-Newman black holes with test particles}, \emph{Phys. Rev.} \textbf{D 87} (2013) 044028.

\bibitem{RS1}
M. Richartz and A. Saa, \emph{Overspinning a nearly extreme black hole and the weak cosmic censorship conjecture¡±}, \emph{Phys. Rev.} \textbf{D 78} (2008) 081503(R).

\bibitem{SH2}
S. Hod, \emph{Weak cosmic censorship: as strong as ever}, \emph{Phys. Rev. Lett.} \textbf{100} (2008) 121101.

\bibitem{BCK1}
E. Barausse, V. Cardoso and G. Khanna, \emph{Test bodies and naked singularities: Is the self-force the cosmic censor?} \emph{Phys. Rev. Lett.} \textbf{105} (2010) 261102.

\bibitem{BCK2}
E. Barausse, V. Cardoso and G. Khanna, \emph{Testing the cosmic censorship conjecture with point particles: The effect of radiation reaction and the self-force}, \emph{Phys. Rev.} \textbf{D 84} (2011) 104006.

\bibitem{ZVPH}
P. Zimmerman, I. Vega, E. Poisson and R. Haas, \emph{Selfforce as a cosmic censor}, \emph{Phys. Rev.} \textbf{D 87} (2013) 041501.

\bibitem{IST}
S. Isoyama, N. Sago and T. Tanaka, \emph{Cosmic censorship in overcharging a Reissner-Nordstrom black hole via charged particle absorption}, \emph{Phys. Rev.} \textbf{D 84} (2011) 124024.

\bibitem{CB}
M. Colleoni and L. Barack, \emph{Overspinning a Kerr black hole: The effect of the self-force}, \emph{Phys. Rev.} \textbf{D 91} (2015) 104024.

\bibitem{CBSM}
M. Colleoni, L. Barack, Abhay G. Shah, and M. van de Meent \emph{Overspinning a Kerr black hole: The effect of the self-force}, \emph{Phys. Rev.} \textbf{D 92} (2015) 084044.

\bibitem{IS}
I. Semiz, \emph{Dyonic Kerr-Newman black holes, complex scalar field and Cosmic Censorship}, \emph{Gen. Relat. Grav.} \textbf{43} (2011) 833.

\bibitem{GZT}
G.Z. Toth, \emph{Test of the weak cosmic censorship conjecture with a charged scalar field and dyonic Kerr-Newman black holes}, \emph{Gen. Relat. Grav.} \textbf{44} (2012) 2019.

\bibitem{DS1}
K. D$\ddot{u}$ztas and I. Semiz, \emph{Cosmic censorship, black holes and integer-spin test fields}, \emph{Phys. Rev.} \textbf{D 88} (2013) 064043.

\bibitem{KD}
K. D$\ddot{u}$ztas, \emph{Overspinning BTZ black holes with test particles and fields}, \emph{Phys. Rev.} \textbf{D 94} (2016) 124031.

\bibitem{DS2}
K. D$\ddot{u}$ztas, \emph{Stability of event horizons against neutrino flux: the classical picture}, \emph{Class. Quantum Grav.} \textbf{32} (2015) 075003.

\bibitem{BG3}
B. Gwak, \emph{Weak cosmic censorship conjecture in Kerr-(Anti-)de Sitter black hole with scalar field}, \emph{JHEP} \textbf{1809} (2018) 081.

\bibitem{RMW2}
R.M. Wald, \emph{Kerr-Newman black holes cannot be over-charged or over-spun,} \emph{Int. J. Mod. Phys.} \textbf{D 27} (2018) 1843003.

\bibitem{SW}
J. Sorce and R.M. Wald, \emph{Gedanken experiments to destroy a black hole. II. Kerr-Newman black holes cannot be overcharged or overspun}, \emph{Phys. Rev.} \textbf{D 96} (2017) 104014.

\bibitem{MST}
Y. Mino, M. Sasaki and T. Tanaka, \emph{Gravitational radiation reaction to a particle motion}, \emph{Phys. Rev.} \textbf{D 55} (1997) 3457.

\bibitem{CO}
J. Crisostomo, R. Olea, \emph{Hamiltonian treatment of the gravitational collapse of thin shells}, \emph{Phys. Rev.} \textbf{D 69} (2004) 104023.

\bibitem{LCNR}
M. Bouhmadi-Lopez, V. Cardoso, A. Nerozzi and J V. Rocha, \emph{Cosmic censorship conjecture in Kerr-Sen black hole}, \emph{Phys. Rev.} \textbf{D 81} (2010) 084051.

\bibitem{RS}
M. Richartz and A. Saa, \emph{Challenging the weak cosmic censorship conjecture with charged quantum particles}, \emph{Phys. Rev.} \textbf{D 84} (2011) 104021.

\bibitem{SH3}
S. Hod, \emph{Cosmic censorship: formation of a shielding horizon around a fragile horizon}, \emph{Phys. Rev.} \textbf{D 87} (2013) 024037.

\bibitem{NQV}
J. Natario, L.Queimada and R. Vicente, \emph{Test fields cannot destroy extremal black holes}, \emph{Class. Quant. Grav.} \textbf{33} (2016) 175002.

\bibitem{HS}
V. Husain and S. Singh, \emph{On the Penrose inequality in anti-de Sitter space}, \emph{Phys. Rev.} \textbf{D 96} (2017) 104055.

\bibitem{RV}
K.S. Revelar and I. Vega, \emph{Overcharging higher-dimensional black holes with point particles}, \emph{Phys. Rev.} \textbf{D 96} (2017) 064010.

\bibitem{BG1}
B. Gwak, \emph{Thermodynamics with pressure and volume under charged particle absorption}, \emph{JHEP} \textbf{1711} (2017) 129.

\bibitem{BG2}
B. Gwak, \emph{Cosmic censorship conjecture in Kerr-Sen black hole}, \emph{Phys. Rev.} \textbf{D 95} (2017) 124050 .

\bibitem{CHS2}
T. Crisford and J.E. Santos, \emph{Violating weak cosmic censorship in $AdS_4$}, \emph{Phys. Rev. Lett.} \textbf{D 118} (2017) 181101 .

\bibitem{GMZZ}
B. Ge, Y. Mo, S. Zhao and J. Zheng, \emph{Higher-dimensional charged black holes cannot be over-charged by gedanken experiments}, \emph{Phys. Lett.} \textbf{B 783} (2018) 440.

\bibitem{KD2}
K. D$\ddot{u}$ztas, \emph{Can test fields destroy the event horizon in the Kerr-Taub-NUT spacetime?} \emph{Class. Quant. Grav.} \textbf{35} (2018) 045008.

\bibitem{KD3}
K. D$\ddot{u}$ztas, \emph{Over-spinning Kerr-Sen black holes with test fields}, \emph{ Int. J. Mod. Phys.} \textbf{D 28} (2018) 1950044.

\bibitem{CHS}
T. Crisford, G.T. Horowitz and J.E. Santos, \emph{Testing the weak gravity-cosmic censorship connection}, \emph{Phys. Rev.} \textbf{D 97} (2018) 066005.

\bibitem{ASZZ}
J. An, J. Shan, H. Zhang, S. Zhao, \emph{Five-dimensional Myers-Perry black holes cannot be overspun in gedanken experiments}, \emph{Phys. Rev.} \textbf{D 97} (2018) 104007.

\bibitem{YW}
T.Y. Yu and W.Y. Wen, \emph{Cosmic censorship and weak gravity conjecture in the Einstein-Maxwell-dilaton theory}, \emph{Phys. Lett.} \textbf{B 781} (2018) 713.

\bibitem{LWL}
B. Liang, S.W. Wei and Y.X. Liu, \emph{Weak cosmic censorship conjecture in Kerr black holes of modified gravity}, [arXiv:1804.06966 [gr-qc]].

\bibitem{DJ}
K. D$\ddot{u}$ztas and M. Jamil, \emph{Testing cosmic censorship conjecture for extremal and near-extremal (2+1)-dimensional MTZ black holes},  [arXiv:1808.04711[gr-qc]].

\bibitem{GG}
Y. Gim and B. Gwak, \emph{Charged black hole in gravity's rainbow: violation of weak cosmic censorship}, [arXiv:1808.05943[gr-qc]].

\bibitem{BG5}
B. Gwak, \emph{Weak cosmic censorship with pressure and volume in charged anti-de Sitter black hole under charged scalar field}, [arXiv:1901.05589[gr-qc]].

\bibitem{KD5}
K. D$\ddot{u}$ztas, \emph{Kerr-Newman black holes can be generically overspun}, \emph{Eur. Phys. J.} \textbf{C 79} (2019) 316.

\bibitem{CHEN}
D.Y. Chen, \emph{Thermodynamics and weak cosmic censorship conjecture in extended phase spaces of anti-de Sitter black holes with particles' absorption}, \emph{Eur. Phys. J.} \textbf{C 79} (2019) 353.

\bibitem{WLM1}
B. Wang, C.Y. Lin and E. Abdalla, \emph{Quasinormal modes of Reissner-Nordstrom anti-de Sitter black holes}, \emph{Phys. Lett.} \textbf{B 481} (2000) 79.

\bibitem{WLM2}
B. Wang, C.Y. Lin and C. Molina, \emph{Quasinormal behavior of massless scalar field perturbation in Reissner-Nordstrom anti-de Sitter spacetimes}, \emph{Phys. Rev.} \textbf{D 70} (2004) 064025.

\bibitem{BK}
E. Berti and K.D. Kokkotas, \emph{Quasinormal modes of Reissner-Nordstrom-anti-de Sitter black holes: scalar, electromagnetic and gravitational perturbations}, \emph{Phys. Rev.} \textbf{D 67} (2003) 064020.

\bibitem{CYZ}
D.Y. Chen, H.T. Yang and X.T. Zu, \emph{Fermion tunneling from anti-de Sitter spaces}, \emph{Eur. Phys. J.} \textbf{C 56} (2008) 119.

\bibitem{PL}
C. Peca and Jose' P.S. Lemos, \emph{Thermodynamics of Reissner-Nordstr$\ddot{o}$m-anti-de Sitter black holes in the grand canonical ensemble}, \emph{Phys. Rev.} \textbf{D 59} (1999) 124007.

\bibitem{MKP1}
Y.S. Myung, Y.W. Kim and Y.J. Park,  \emph{Ruppeiner geometry and 2D dilaton gravity in the thermodynamics of black holes}, \emph{Phys. Lett.} \textbf{B 663} (2008) 342.

\bibitem{MKP2}
Y.S. Myung, \emph{Phase transition between non-extremal and extremal Reissner-Nordstrom black holes}, \emph{Mod. Phys. Lett.} \textbf{A 23} (2008) 667.

\bibitem{NTW}
C. Niu, Y. Tian and X. Wu, \emph{Critical phenomena and thermodynamic geometry of RN-AdS black holes}, \emph{Phys. Rev.} \textbf{D 85} (2012) 024017.

\bibitem{BGA}
B. Ganchev, \emph{Superradiant instability in AdS}, [arXiv:1608.01798 [hep-th]].







\end{thebibliography}
\end{document}